# ONE-DIMENSIONAL TRAFFIC FLOW MODELS: THEORY AND COMPUTER SIMULATIONS*

J.G. Brankov, N.C. Pesheva, N.Zh. Bunzarova
Institute of Mechanics, Bulg. Acad. Sci., Acad. G. Bonchev St., Block 4, 1113 Sofia, Bulgaria
e-mails: brankov@imbm.bas.bg, nina@imbm.bas.bg, nadezhda@imbm.bas.bg

**Abstract** Theoretical advances in the study of non-equilibrium phenomena are briefly reviewed with emphasis on steady state properties of one-dimensional driven lattice gases. The presentation is focused on the totally asymmetric simple-exclusion process (TASEP) with open boundary conditions: particles are injected at the left end with rate $\alpha$ and removed at the right end with rate $\beta$. Depending on the values of these parameters, the model exhibits three stationary phases, separated by lines of first- and second-order non-equilibrium phase transitions. New simulation results on the power spectrum of the fluctuating total number of particles in the different phases of the system are presented. Our theoretical contribution concerns the approximate evaluation of the power spectrum in the domain-wall picture of the coexisting low- and high-density phases. Finally, we review some of our recent results on the TASEP defined on an open network containing a double-chain section in the middle. With the aid of a simple theory, which neglects correlations at the junctions of the chain segments, the possible phase structures of the model are found. Density profiles and nearest-neighbor correlations in the steady states of the model at representative points of the phase diagram are obtained by means of computer simulations. On the coexistence line cross-correlations are found to exist between equivalent sites in the branches of the middle section.

## 1. Introduction: Non-equilibrium Phenomena

Non-equilibrium phenomena take place either when the system is relaxing towards an equilibrium state, or when it is maintained away from equilibrium by external forces. Presently, the following features are considered as characteristic of the complex non-equilibrium behavior: phase transitions between stationary states of one-dimensional systems driven by boundary conditions; spatiotemporal pattern formation, including surface growth and density waves; generic long-range correlations due to conservation of current; anomalous diffusion. Here we will focus on a special class of simple lattice-gas models driven far from equilibrium - the so called driven diffusive systems, introduced by Katz, Lebowitz and Spohn [1]. In these models the particle dynamics is specified by stochastic hopping rules to nearest vacant sites with biased rates that simulate the action of an external field or a thermodynamic force. For example, such a situation takes place for charged particles in external electric field, or by coupling the opposite ends of a system of neutral particles to different thermal baths or particle reservoirs.

The fundamental understanding and theoretical classification of non-equilibrium steady states is incomplete. It is known that under constant drive the initial state of the system evolves with time towards a steady state with stationary current of particles. Many theoretical and numerical results have been obtained, but the structure of the non-equilibrium steady states is still not well understood. The density fluctuations of a conserved (in the bulk) quantity, such as the local particle density, were found to decay algebraically over distances comparable with the system size. Naturally, in the presence of long-range correlations, the boundary effects turn out to be essential. Steady states have been described exactly and studied in detail just in a few exceptional cases of one-dimensional (1D) models. For recent reviews on exactly solvable models of stochastic particle systems far from equilibrium we refer the reader to [2,3].

---

## 2. Model and Applications

Exactly solved models of driven diffusive systems can be obtained by switching off the inter-particle interaction, leaving only the hard-core on-site exclusion and the external drive. Thus one arrives at the asymmetric simple exclusion process (ASEP) which has been extensively studied on simple chains with periodic, closed and open boundary conditions. The ASEP is one of the simplest models of driven many-particle systems with particle conserving stochastic dynamics. In the extremely asymmetric case particles are allowed to move, with probability $p$, in one direction only - this is the totally asymmetric simple exclusion process (TASEP), illustrated in Fig. 1. It was first introduced in [4] as a model of protein synthesis. For its description in the context of interacting Markov processes see [5]. For example, TASEP and its generalizations serve as prototype models of: kinetics of protein synthesis [4] and biological transport [13,14]; fast ionic conductors [15]; vehicular traffic flow [16-18]; one-dimensional surface growth [19-21]; diffusional pumping in zeolites and biomembrane channels [22]; transport of 'data packets' in the Internet [23]; limit order market [24].

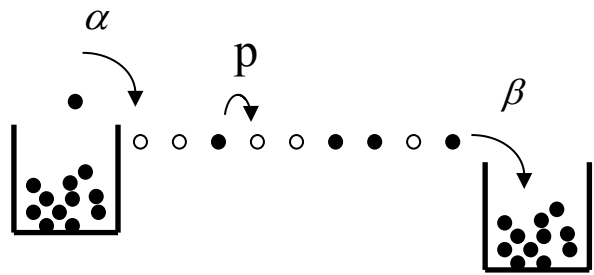

**Figure 1.** Schematic representation of the one-dimensional TASEP

The steady states of TASEP are exactly known for both open and periodic boundary conditions, for continuous-time and several kinds of discrete-time dynamics. The continuous-time dynamics is modeled by the so called random-sequential update: the algorithm chooses with equal probability any one of the lattice sites (the left reservoir is included as an additional site), and, if the chosen site is occupied by a particle, moves it (with $p = 1$) to the nearest-neighbour site on the right, provided the target site is empty. In the case of open systems, particles are injected at the left end with rate $\alpha$ and removed at the right end with rate $\beta$. When $\alpha, \beta \in (0,1]$, the boundary conditions correspond to coupling of the system to reservoirs of particles with constant densities $\alpha$ and $1-\beta$, respectively. The exact stationary states have been found by the Matrix Product Ansatz (MPA) in [6]. The MPA formalism has been extended to the TASEP with discrete-time dynamics: sublattice parallel [7], sequentially ordered [8,9], and fully parallel [10,12]. As predicted by Krug [12], the change of the boundary rates induces non-equilibrium phase transitions between different stationary phases. In the thermodynamic limit, the phase diagram of the stationary states in the plane of the particle injection and removal rates is shown in Fig. 2. It exhibits three distinct phases: a low-density free-flow phase (region $AI \cup AII$), a high-density congested traffic one (region $BI \cup BII$), and a maximum current phase (region $MC$), characterized by a synchronized flow in which jams and free-flow coexist at intermediate densities. These phases are separated by lines of non-equilibrium first-order and second-order phase transitions. For many other traffic models it is still an open question whether the transition from one stationary regime to another takes place as a true phase transition, or is a smooth crossover.

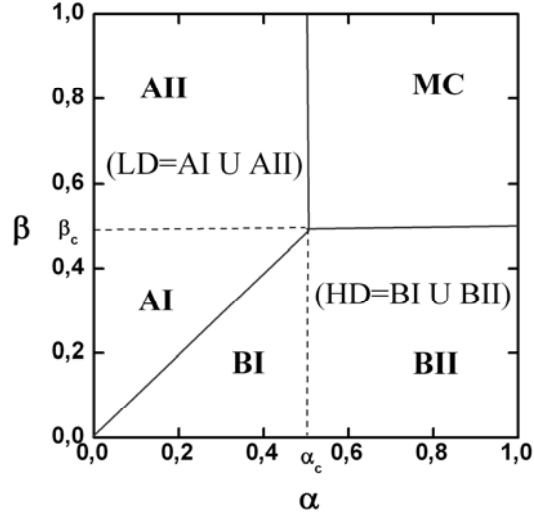

**Figure 2.** The phase diagram of the one-dimensional TASEP as a function of the injection and removal rates, $\alpha$ and $\beta$, respectively, exhibits low density (LD), high density (HD) and maximal current (MC) phases.

## 3. Theory of Dynamic Correlations

Among the semi-phenomenological though very successful analytical approaches to the dynamic properties of the TASEP we outline the one based on Boltzmann-Langevin type equations. In its framework the Boltzmann kinetic equation is supplied with a Langevin noise term which takes into account the stochastic nature of the particle dynamics. This approach was introduced in condensed matter physics by Kogan and Shul'man [25] and it is believed to reproduce correctly the properties at long-time and large-distance scales. Here, following [26], we outline its application to the TASEP. The particle configurations of the TASEP on a chain of $L$ sites are described by a set of binary random variables $\{\tau_i(t)\}_{i=1}^{L}$ where $\tau_i(t)=1$ ($\tau_i(t)=0$) means occupied (vacant) lattice site $i \in \{1,2,\ldots,L\}$ at microscopic time $t$. Let $j_i(t) = \tau_i(t)[1-\tau_{i+1}(t)]$ be the random current through the bond $(i, i+1)$ at time $t$. Then the lattice continuity equation has the form

$$\partial_t \tau_i(t) = j_{i-1}(t) - j_i(t). \tag{1}$$

The right-hand side of this equation yields the balance of in-coming and out-going currents and thus resembles the Boltzmann collision intergal. To pass to the continuum limit one fixes the total length of the system, say, $La = 1$, where $a$ is the lattice spacing, and considers the limit $L \to \infty$, hence $a = 1/L \to 0$. To account for the fluctuation effects in a stationary state, one separates the deterministic and fluctuating parts of the local density and current, assuming constant in time and uniform in space (for the sake of simplicity) bulk density $\bar{\rho} = \text{const}$:

$$\tau_i(t) \to \bar{\rho} + \delta\rho_i(t), \quad j_i(t) \to [\bar{\rho} + \delta\rho_i(t)][1 - \bar{\rho} - \delta\rho_{i+1}(t)] + \delta j_i(t). \tag{2}$$

Next, one introduces a macroscopic spatial coordinate $x = ia$ and makes a Taylor expansion of the local density fluctuations to the second order in $a$:

$$\delta\rho_{i\pm 1} = \phi(x,t) \pm a\partial_x \phi(x,t) + \frac{1}{2}a^2 \partial_x^2 \phi(x,t), \tag{3}$$

where $\phi(x,t)$ is the density fluctuation field. The current fluctuations reflect the stochastic particle dynamics and are described in the lowest-order gradient expansion as [26]:

$$\delta j_{i-1}(t) - \delta j_i(t) \simeq -a\partial_x \eta(x,t), \tag{4}$$

where $\eta(x,t)$ is uncorrelated at different space-time points Gaussian noise field with zero mean value and density-dependent variance $\Delta = J(\bar{\rho})[1-J(\bar{\rho})]$. Note that here $J(\rho) = \rho(1-\rho)$ is the mean-field approximation for the current in a stationary state with uniform density $\rho$. Finally, after rescaling of time, $t \to t/a$, one arrives at the stochastic nonlinear evolution equation with infinitesimal (as $a \to 0$) diffusive term

$$\partial_t \phi(x,t) + [(1-2\bar{\rho}) - 2\phi(x,t)]\partial_x \phi(x,t) = \frac{1}{2}a\partial_x^2 \phi(x,t) - \partial_x \eta(x,t). \tag{5}$$

This equation is valid away from the phase boundary $\alpha = \beta < 1/2$, where the average density profile can be considered as uniform. The quantity $\text{v} = 1 - 2\bar{\rho}$ represents the collective velocity of density perturbations and obeys the exact non-equilibrium fluctuation-dissipation theorem $\text{v}(\rho) = \partial_\rho J(\rho)$. The collective velocity v changes sign at $\bar{\rho} = 1/2$, when the current attains its maximal value $J = 1/4$. It is seen from eq. (5) that the nonlinear term, proportional to $\phi \partial_x \phi$, is essential only for small values of $v$, i.e. close to the coexistence line $\alpha = \beta < 1/2$ of the low- and high density phases, and close to the boundaries with the maximum current phase. For densities $\bar{\rho} \neq 1/2$, one can neglect the nonlinear term and obtain an explicit expression for the density-density correlation function from the resulting linearized Boltzmann-Langevin equation:

$$C_{LBL}(x,t) = \langle \phi(x,t)\phi(0,0) \rangle = \frac{\Delta}{\sqrt{2\pi |t|}} \exp\left[-\frac{(x-\text{v}t)^2}{2|t|}\right]. \tag{6}$$

At $\bar{\rho} = 1/2$ eq. (5) becomes exactly the noisy Burgers equation which belongs to the universality class of the Kardar-Parisi-Zhang (KPZ) equation [24]. Universality is understood as robustness of the exponents describing the large-scale power-law behaviour with respect to changes of details such as inter-particle interactions (provided they remain short-ranged) and microscopic dynamics. The KPZ equation describes non-equilibrium wetting (layer growth) on top of a hard substrate. It is written for the height $h(x,t)$ of the layer above the point $x$ at time $t$, and the gradient $\phi(x,t) = \partial_x h(x,t)$ satisfies the noisy Burgers equation. For the KPZ universality class it is known that at large spatial and temporal separations the pair correlation function obeys the scaling form:

$$C_{KPZ}(x,t) \cong x^{-1} F(t/x^{3/2}). \tag{7}$$

Here the scaling function $F$ has the asymptotic behavior $F(y) \propto y^{-2/3}$ as $y \to \infty$.

In the cases when the average density profile is not flat, one may use the phenomenological domain wall (DW) theory [28]. Its main idea is that, when $\alpha \neq 1-\beta$, each reservoir tends to enforce a domain in the chain with its own density: the left domain with density $\rho_- = \alpha$, and the right one with density $\rho_+ = 1-\beta$. At a given time $t$ these domains may coexist, being separated by a domain wall. On the deterministic level of approximation, by neglecting the current and density fluctuations and correlations, one obtains the following nonlinear evolution equation for the average local density $\rho(x,t)$ (dependent on the rescaled temporal and spatial coordinates):

$$\partial_t \rho + (1-2\rho)\partial_x \rho = \frac{1}{2}a\partial_x^2 \rho. \tag{8}$$

For an open system, coupled at the left and right ends to reservoirs of particles with constant densities $\alpha$ and $1-\beta$, respectively, this equation has to be supplied with the boundary conditions:

$$\rho(0,t) = \alpha, \quad \rho(1,t) = 1-\beta. \tag{9}$$

On the coexistence line, $0 < \alpha = \beta < 1/2$, and for small values of the lattice spacing $a > 0$, its stationary solution

$$\rho_0(x) \cong \frac{1}{2} + \frac{1-2\alpha}{2}\tanh\left[\frac{1-2\alpha}{a}(x-x_0)\right] \tag{10}$$

describes a smooth domain wall between the low-density and high-density phases, which can be located with equal probability at arbitrary point $x_0 \in (0,1)$. The shape (10) of the domain wall is valid away from the boundaries: $x_0/a \gg 1$ and $(1-x_0)/a \gg 1$.

Upon shifting from the coexistence line, for small $a > 0$ and in a limited interval of times, $0 < t < 1/|\beta - \alpha|$, eq. (8) has a traveling-wave solution of the form

$$\rho(x) \cong \frac{1-V}{2} + \frac{u}{2}\tanh\left[\frac{u}{a}(x-Vt)\right], \tag{11}$$

with amplitude $u = 1 - \alpha - \beta$ and constant speed $V = \beta - \alpha$. Thus, when $\alpha < \beta < 1/2$ the domain wall moves to the right until it reaches the end of the system and the stationary low-density phase is established. In the opposite case, $\beta < \alpha < 1/2$, the domain wall travels to the left boundary of the system and, after sticking to it, the stationary high-density phase takes place. Note that in the continuum limit $a \to 0$ the domain wall becomes a sharp interface between two constant densities: $\rho_- = \alpha$, and $\rho_+ = 1 - \beta$.

The above deterministic picture has to be complemented with a stochastic element. Indeed, each time a particle enters the system, the domain wall moves to the left, and each time a particle leaves the system it moves to the right. Thus it performs a random walk on the chain. On the coexistence line $\alpha = \beta < 1/2$, we have $V = 0$, and the random walk is symmetric, with diffusion constant $D = \alpha(1-\alpha)/(1-2\alpha)$ and reflecting boundary conditions. This diffusive motion accounts for the power spectrum of the particle density fluctuations at small frequencies. In addition to the fluctuations of the domain wall position, the local density on both sides of the domain wall fluctuates according to the predictions of the linearized Boltzmann-Langevin equation for the corresponding low- and high-density phases. Both of these effects contribute to the density-density correlation function. Since they take place on different time scales, as a reasonable approximation one can assume additivity of the corresponding contributions.

Finally, we mention that away from the coexistence line, the domain wall performs a biased random walk with different hopping rates $D_r = J_+/(\rho_+ - \rho_-)$ and $D_l = J_-/(\rho_+ - \rho_-)$, to the right and left, respectively. Here $J_-$ and $J_+$ are the currents of particles characteristic of each of the two domains, $J_\pm = \rho_\pm(1-\rho_\pm)$.

## 4. Fluctuation Spectra of TASEP on a Chain

Consider the fluctuations of the random total number $N(t)$ of particles on a finite open chain:

$$S_N(\omega) = \lim_{T \to \infty} \frac{1}{2T} \left\langle \left| \int_{-T}^{T} dt\, N(t) e^{i\omega t} \right|^2 \right\rangle_{st}. \tag{12}$$

According to the Wiener-Khinchin theorem, if [33]

$$\lim_{T \to \infty} \frac{1}{2T} \int_{-T}^{T} d\tau\, |\tau|\, e^{i\omega\tau} C_N(|\tau|) = 0, \tag{13}$$

then an equivalent definition of the power spectrum can be given in terms of the autocorrelation function:

$$S_N(\omega) = \int_{-\infty}^{\infty} d\tau\, e^{i\omega\tau} C_N(|\tau|). \tag{14}$$

In section 4.1 we present our direct simulation results on the power spectrum $S_N(2\pi f)$ of the fluctuating total number of particles $N(t)$ in the open TASEP with continuous-time dynamics, as a function of the frequency $f = \omega/2\pi$, in different regions of the phase diagram. The interest in the low-frequency behaviour of $S_N(\omega)$ has arisen due to the attempts to find a general theoretical explanation of the so-called $1/f$-noise in the framework of the concept of self-organized criticality (SOC) [29]. The idea that driven lattice gas models with deterministic bulk dynamics can exhibit SOC with $1/f$ power spectrum was advanced by Jensen [30,31], who estimated from direct numerical results $S_N(\omega) \propto \omega^{-a}$, with $a \approx 1.5$ for systems in one dimension and $a \approx 1.2$ in two dimensions. His explanation of the $1/f$ noise [32], based on the linear diffusion equation with boundary noise term, was criticized in [33]. However, the value $a \approx 1.5$ has been obtained by Andersen et. al. [34] also for lattice gas models with stochastic MC dynamics in one, two, and three dimensions. This result is characteristic of non-interacting random walks. The correct exponents were found by Krug [12] on the basis of the Boltzmann-Langevin type equation with noise, eq. (5): by neglecting the boundary conditions on the density fluctuation field $\phi(x,t)$, he predicted the exponent $a = 2$ in the case of $1 - 2\bar{\rho} \neq 0$ (in the low- and high-density phases), and $a = 5/3$ when $\bar{\rho} = 1/2$ (in the maximum current phase). More recently, it has been claimed that a signature of $1/f$-noise was found in the stochastic traffic model of Nagel and Schreckenberg at the critical density of the jamming transition [35]. For experimental investigations of SOC in real traffic see [36]. By now, most of the low-frequency behaviour of the power spectrum of the number of particles in the open TASEP has already found its theoretical explanation. In section 4.2, we add to the existing knowledge our analytic evaluation of $S_N(\omega)$ at the coexistence line, $\alpha = \beta < 1/2$, in the framework of the domain wall theory.

**4.1. Numerical simulations of the power spectrum**

We start by noting that computer simulations result in a finite set of data: $\{N(t), t = 1, 2, ..., T\}$. The power spectrum of the total number of particles is then defined as:

$$S_{N,T}(\omega) = \frac{1}{T}\left\langle \left|\hat{N}_T(\omega)\right|^2 \right\rangle, \quad \text{where} \quad \hat{N}_T(\omega) = \sum_{t=1}^{T} e^{i\omega t} N(t). \tag{15}$$

To the finite set of data points there corresponds a finite set of $T$ independent Fourier transforms $\hat{N}_T(\omega_n)$ corresponding to the frequencies $\omega_n = 2\pi n/T$, $n = 1, 2, ..., T$, or, equivalently, $n = 0, 1, ..., T-1$. The simulations start from a random initial configuration with the wanted density and the system is let to evolve with time according to the random sequential update of configurations. Recording of the necessary data begins after a sufficiently long period of time. It has been analytically found that the relaxation time $\tau_r$ to the stationary state of the TASEP under periodic boundary conditions diverges with the system size $L$ as $\tau_r \propto L^z$, with dynamic exponent $z = 3/2$. For open boundary conditions exact results are not available, but some renormalization-group studies indicate that the relaxation times are finite in both the low- and high-density phases, while on their coexistence line and everywhere in the maximum current phase $\tau_r \propto L^{3/2}$.

Our computer simulations results, shown in Figs. 3 and 4, reveal different types of fluctuation behaviour in the different regions of the phase diagram, thus adding new insight to the very intriguing behavior of the TASEP. In the maximal current phase, see Fig. 4, the fluctuations of $N(t)$ are showing non-trivial behaviour characteristic of SOC.

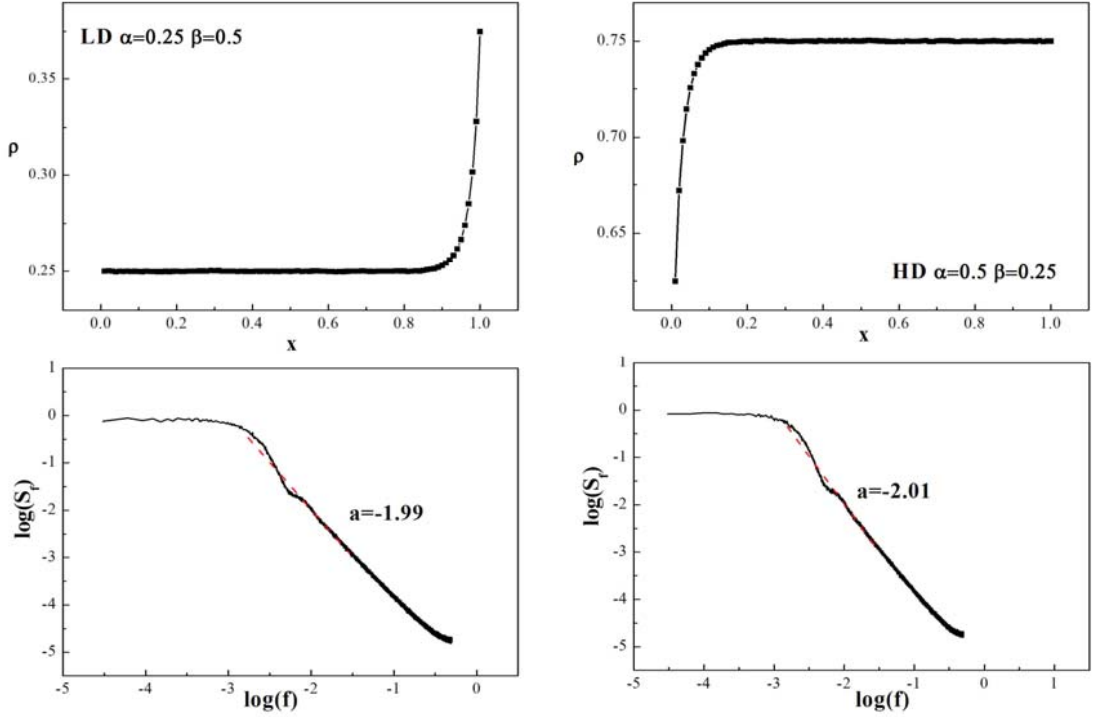

**Figure 3.** Simulation results ($L$=100): the density profiles (upper graphs) and the corresponding power spectra (lower graphs) of the particle numbers in the LD and HD phases. In the LD phase the exponent of the power spectrum varies with the density $\rho$, it decreases from $a \cong 2$ at very low densities, to $a \cong 1.66$ as the density reaches $\rho = 1/2$. Similarly in the HD phase the exponent $a$ decreases with $(1-\rho)$.

We find that in the whole MC phase the power spectrum of the total number of particles $N(t)$ is characterized by the same exponent $a \cong 1.66$. This result is consistent with the findings in [12]. Across the CL the exponent $a$ changes quite sharply from $a \cong 1.66$ (at $\alpha = \beta = 1/2$) to $a \cong 2$ (in the LD and HD phases). The power spectrum of the total number of particles $N(t)$ is very well described by our simple domain wall theory at small values of $\alpha$, the deviation grows for higher values of $\alpha < 1/2$. The latter deviation can be explained by the increasing with $\alpha$ contribution of the fluctuations in the pure low- and high-density phases on both sides of the domain wall.

**4.2. Power spectrum in the domain wall picture**

The fluctuations induced by the Brownian motion of the domain wall between the coexisting low-density and high-density phases of the TASEP have been studied recently in [37,38]. In [37] the single-time variance of the total number of particles in the system has been obtained in a dynamical regime and compared with direct simulation results. The authors of [38] have investigated the power spectrum $I(\omega)$ of the local density fluctuations and found that in a certain range of low-frequencies $\omega$ it obeys the power law $I(\omega) \propto \omega^{-3/2}$.

Here we present our analytic evaluation of the power spectrum $S_N(\omega)$ of the fluctuation of the total number of particles $N(t)$ on a open chain of $L$ sites, with continuous-time dynamics, at the coexistence line $\alpha = \beta < 1/2$. We adopt an idealized model of a sharp domain wall between the phases with average low density, $\rho_- = \alpha < 1/2$, and high density, $\rho_+ = 1 - \alpha > 1/2$, similar to the one used in [37,38]. The random position $\xi$ of the domain is allowed to take $L+1$ values, identified with the sites $n = 0,1,...,L$ of an auxiliary chain.

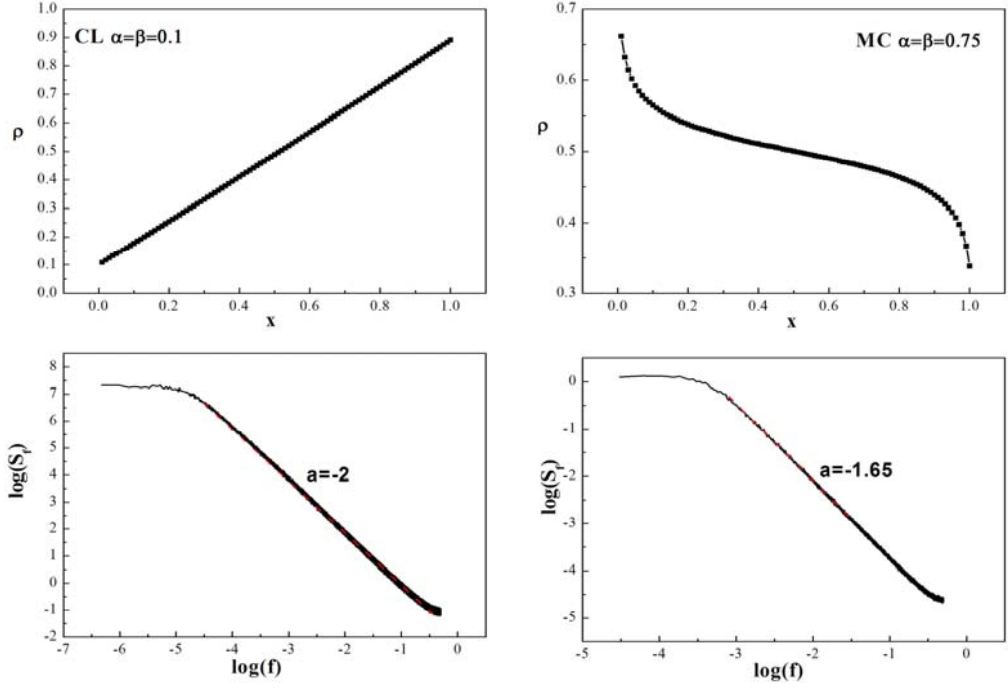

**Figure 4.** Simulation results ($L$=100): the density profiles (upper graphs) and the corresponding power spectra (lower graphs) of the particle numbers on the coexistence line (CL) and in the maximal current (MC) phase.

In accordance with the domain wall picture, reviewed in section 3, we assume that the wall performs a symmetric continuous-time random walk. At each moment of time $t$, the random total number of particles in the TASEP on a chain of $L$ sites is given by $N(t) = \rho_+ L - (\rho_+ - \rho_-)\xi_t$, where $\xi_t \in \{0,1,...,L\}$ is the momentary random position of the wall. The stationary state probability distribution of $\xi$ is uniform, i.e., $P_{st}(\xi = n) = 1/(L+1)$ for all $n = 0,1,...,L$. Hence, denoting by $<\zeta>_{st}$ the average value of a random variable $\zeta$ in a stationary state, we have $<\xi_t>_{st} = L/2$ and $<N(t)>_{st} = L(\rho_+ + \rho_-)/2 = L/2$. The random deviations of the total number of particles from the stationary average value are represented by the variable $\delta N(t) = N(t) - <N(t)>_{st} = \Delta(L/2 - \xi_t)$, where $\Delta = 1 - 2\alpha$. The central object of our study is the time dependent correlation function:

$$C_N(|t-t'|) \equiv <\delta N(t)\delta N(t')>_{st} = \Delta^2(<\xi_t\xi_{t'}>_{st} - L^2/4), \qquad (16)$$

and its finite-time Fourier transform:

$$S_{N,T}(\omega) = C_N(0) + 2\sum_{\tau=1}^{T-1} \cos(\omega\tau)C_N(\tau). \qquad (17)$$

The explicit expression for the correlation function in the right-hand side of eq. (16) reads:

$$<\xi_t\xi_{t'}>_{st} = \sum_{n_0=0}^{L} P_{st}(n_0) \sum_{n'=0}^{L} \sum_{n=0}^{L} nn' W(n_0,0 \to n',t') W(n',t' \to n,t), \qquad (18)$$

where $W(n',t' \to n,t)$ is the transition rate from position $n'$ at time $t'$ to position $n$ at time $t$, and the initial probability distribution is the stationary one $P_{st}(n_0) = 1/(L+1)$. The calculations are performed with the aid of the transition rates for a random walk on the chain $n = 0,1,...,L$ with reflecting boundary conditions at $n = 0$ and $n = L$, known from [39], see also [37,38]:

$$W(n',t' \to n,t) = \frac{1}{L+1}\left\{1 + \sum_{r=1}^{L+1}\left[\cos\frac{\pi(n+n'+1)r}{L+1} + \cos\frac{\pi(n-n')r}{L+1}\right]\exp\left[-2tD\left(1-\cos\frac{\pi r}{L+1}\right)\right]\right\}, \quad (19)$$

where $D = \alpha(1-\alpha)/(1-2\alpha)$ is the diffusion constant, see section 3. To simplify the notation, we set in the remainder

$$s_l^2 \equiv \sin^2\frac{\pi(2l+1)}{2(L+1)} . \quad (21)$$

The result for the correlation function (18) is

$$C_N(|t-t'|) = \frac{\Delta^2}{2(L+1)^2}\sum_{l=0}^{[L/2]}\exp\left[-4|t-t'|D\sin^2\frac{\pi(2l+1)}{2(L+1)}\right]\frac{1-s_l^2}{s_l^4} . \quad (20)$$

Next, we notice that $s_l^2 \geq s_0^2$ for all $l = 0,1,...,[L/2]$, and consider the case of sufficiently long time series, when $T/L^2 \gg 1$, so that in the expression for $S_{N,T}(\omega)$ one can neglect terms of the order $\exp(-4s_l^2 T)$. By using this approximation, which amounts to taking the limit $T \to \infty$, we obtain the power spectrum in the form:

$$S_{N,\infty}(\omega) = \frac{\Delta^2}{2(L+1)^2}\sum_{l=0}^{[L/2]}\frac{1-s_l^2}{s_l^4}\frac{1-\exp(-8Ds_l^2)}{1-2\exp(-4Ds_l^2)\cos(\omega)+\exp(-8Ds_l^2)} . \quad (22)$$

The large-$L$ and small-$\omega$ asymptotic behaviour of the power spectrum, which we are interested in, is determined by the small-$l$ terms in the sum in the right-hand side of the above expression. Therefore, we may expand in terms of $s_l^2$, $\omega$, $L^{-1}$, and extend to infinity the upper limit of summation, to obtain the approximation:

$$S_{N,\infty}(\omega) \cong \frac{16D\Delta^2}{\pi^2}\sum_{l=0}^{\infty}\frac{1}{(2l+1)^2[D^2\pi^4(2l+1)^4/L^4+\omega^2]} . \quad (23)$$

The following features of the power spectrum can be inferred from expression (23):

**1.** There is a crossover frequency, $\omega_c = D\pi^2/L^2$, such that for $\omega \ll \omega_c$, the power spectrum saturates at the value $S_{N,\infty}(0) \cong \Delta^2 L^4/(60D)$. This saturation value reflects the boundedness of $N(t)$ on a finite lattice of $L$ sites.

**2.** In the asymptotic interval $\omega_c \ll \omega \ll 1$ the power spectrum exhibits the power-law behaviour $S_{N,\infty}(\omega) \cong (2D\Delta^2)\omega^{-2}$, which is characteristic of the Brownian motion of the wall.

The above rather crude approximation yields fairly good quantitative explanation of the low-frequency features of the simulated spectra, see Fig. 4, especially for smaller values of $\alpha = \beta < 1/2$. Thus, for $L=100$, and $\alpha = 0.1, 0.2, 0.3, 0.4$, we estimate from the domain wall (DW) theory $f_c^{\text{DW}} \cong$ $1.77 \times 10^{-5}$, $4.19 \times 10^{-5}$, $8.25 \times 10^{-5}$, $1.88 \times 10^{-4}$, and, after rescaling to the same normalization, $\log S_{N,\infty}^{\text{DW}}(0) \cong 7.28, 6.65, 6.01, 5.05$, respectively, while the simulation results yield $f_c^{\text{sim}} \cong$ $2 \times 10^{-5}$, $5 \times 10^{-5}$, $8 \times 10^{-5}$, $1.6 \times 10^{-4}$, and $\log S_{N,T}^{\text{sim}}(0) \cong 7.23, 6.66, 6.09, 5.4$, respectively. The white-noise level and the crossover frequency were evaluated by fitting the power spectrum with the model function suggested in [40]. Obviously, the noticeable difference $S_{N,T}^{\text{sim}}(0) - S_{N,\infty}^{\text{DW}}(0)$, appearing at larger values of $\alpha = \beta \geq 0.3$, has to be attributed to the particle number fluctuations in the pure low-density and high-density phases on both sides of the domain wall.

## 5. Stationary States of a Network

One of the natural physical interpretations of the totally asymmetric version of TASEP is given in terms of a single-lane vehicular traffic, see the recent review [18]. The fully parallel dynamics is considered as most appropriate for traffic modelling, and it is layed in the basis of more sophisticated update rules [16-18, 41]. We mention that the TASEP with parallel update results in the Nagel-Schreckenberg model [16] with maximum vehicle velocity $v_{max} = 1$. Recently much attention has been paid to cellular automaton models of traffic on roads with localized inhomogeneities modelling on- and off-ramps [42-44], and a roundabout at the intersection of two perpendicular streets [45]. Such spatial inhomogeneities were shown to lead to different dynamical phases of congested traffic.

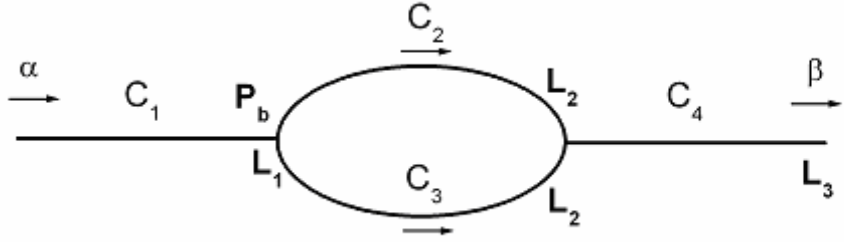

**Figure 5.** Schematic representation of the network: a single lane with two-chain section in the middle. All the segments $C_i$ have equal length $L$. The particles are injected at the left end with a rate $\alpha$ and removed at the right end with a rate $\beta$. The particles move from left to right, at the branching point $L_1$ they take with equal probability $P_b = 1/2$ the upper or the lower branch.

Here we present results of our study of TASEP on a graph consisting of a simple chain with a loop in the middle [46]. A schematic representation of the model is shown in Fig. 5. The last site $i = L_1$ of the head chain is a branching point from which the particles can take the upper or the lower branch of the loop with equal probability $P_b = 1/2$. Simultaneous and independent traffic of particles on the two branches was simulated.

Let us denote the phase structure of the model by $(X_1, X_{2,3}, X_4)$, where $X_i \in (LD, HD, MC, CL)$ denotes the stationary phase of the chain segment $C_i$, $i = 1, 2, 3, 4$. We remind the reader our abbreviations: LD – low density, HD – high density, MC – maximum current, CL – coexistence line. Our analytical analysis of the allowed phase structures, based on the neglect of the pair correlations between the nearest-neighbor occupation numbers belonging to different chain segments, yields the 8 possibilities given below:

(LD,LD,LD) if $\alpha < 1/2$ and $a < \beta$, (24)
(HD,HD,HD) if $\beta < 1/2$ and $a > \beta$, (25)
(LD,LD,HD) or (LD,HD,HD) or (LD,CL,HD) if $\alpha = \beta < 1/2$, (26)
(MC,LD,MC) or (MC,LD,MC) or (MC,CL,MC) if $\alpha > 1/2, \beta > 1/2$. (27)

However, our computer simulations, presented in Figs. 6-13, show that whenever the chain segments $C_{2,3}$ of the middle section may exist in either the low- or high-density phases, they are always found on the coexistence line. The main results of our computer simulations are obtained for chain segments of equal number of sites $L = 100, 200$ (total length of $3L+1$). Time is measured in units of $(3L+1)$ local trials respectively, which we call steps per site (SPS). Below we present the results for the different phase structures observed. Since cases (24) and (25) are very similar, we show here only the results for case (25).

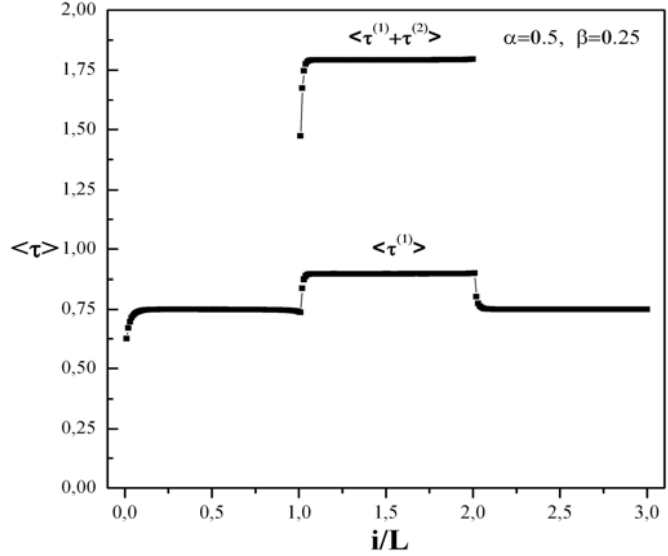

**Figure 6.** Simulation results: Typical density profiles $\langle \tau \rangle$ as function of the scaled distance $i/L$ in the (HD,HD,HD) state at $\alpha = 0.5$, $\beta = 0.25$.

**Case** $\beta < 1/2$ and $\alpha > \beta$ (Figs. 6,7). As expected from our preliminary analysis, see Eq. (25), the phase structure (HD,HD,HD), or respectively (LD,LD,LD) in case (24), is realized, see Fig. 6. The comparison of the theoretical predictions and the simulation results for the current and the bulk densities in the simple-chain segments shows that they agree fairly well, within expected finite-size and finite-sample corrections. Even details of the shape of the density profiles are well explained. Short-range correlations appear in the neighborhood of the bifurcation and merging sites of the network as can be seen in Fig. 7.

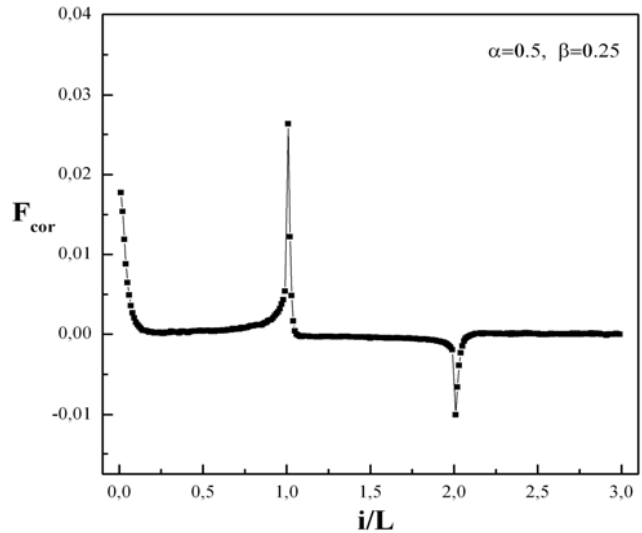

**Figure 7.** Simulation results: Nearest-neighbor correlations $F_{cor}$ in the (HD,HD,HD) state at $\alpha = 0.5$, $\beta = 0.25$.

Otherwise, the properties of the simple-chain segments are close to those expected on the grounds of the approximation which ignores the above mentioned correlations.

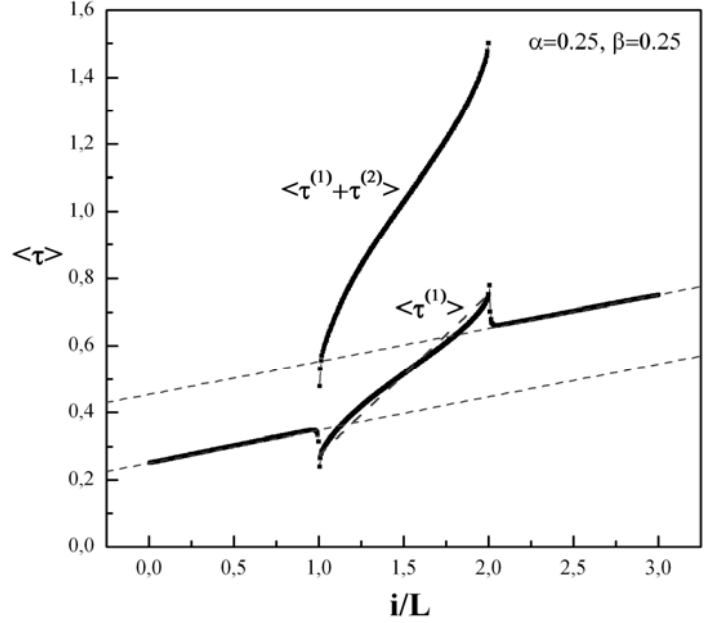

**Figure 8.** Simulation results: Typical density profiles $\langle \tau \rangle$ as function of the scaled distance $i/L$ in the (LD,CL,HD) phase structure at $\alpha = 0.25$, $\beta = 0.25$.

**Case** $\alpha = \beta < 1/2$ (Figs. 8, 9). Most surprising are the results of our computer simulations in this case, see Fig. 8. Since we have no control over the local densities at the bifurcation and merging points, we cannot force the system in the phase structures, given by Eq. (26). Therefore we expect to find the chain segments of the middle section on the coexistence line. Unexpected feature revealed by the simulations is the existence of a non-vanishing slope in the density profiles of the head and tail segments, which are expected to be in the low- and high-density phase, respectively. An inspection of the nearest-neighbor correlations shown in Fig. 8 reveals two important features.

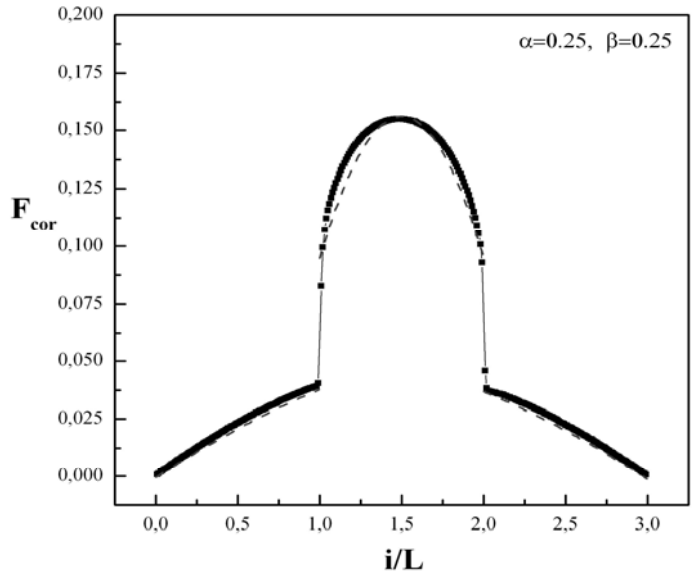

**Figure 9**. Simulation results: Nearest-neighbor correlations $F_{\text{cor}}$ in the (LD,CL,HD) state at $\alpha = 0.25$, $\beta = 0.25$.

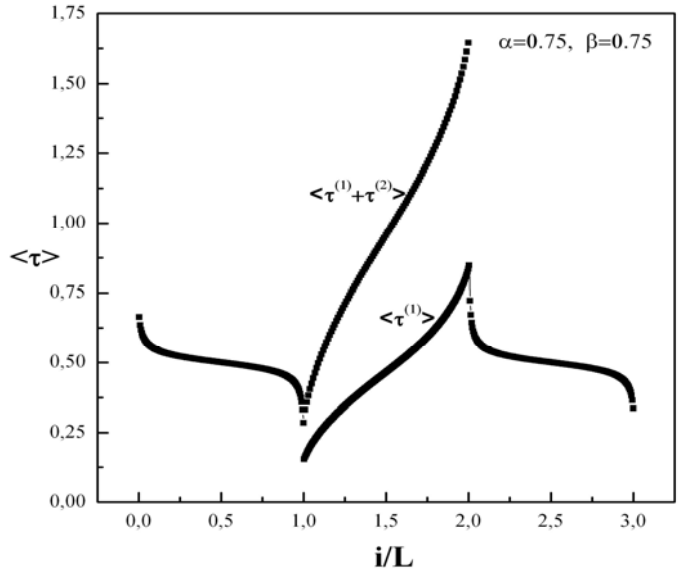

**Figure 10.** Simulation results: Density profiles $\langle\tau\rangle$ as function of the scaled distance $i/L$ with the (MC,CL,MC) phase structure at $\alpha = 0.75$, $\beta = 0.75$.

First, strong correlations with parabolic spatial dependence in the chain segments of the middle section, which is indicative of phase coexistence with completely delocalized domain wall between the low- and high-density phases. Second, rather strong correlations developing in the head and tail chain segments away from the open boundaries. Our simple theory [46] accounts quite well for these results.

**Case** $\alpha > 1/2$ *and* $\beta > 1/2$ (Figs. 10-12). In this case the simulations show that the (MC,CL,MC) phase structure is realized out of the three possibilities, given by Eq. (27). The local density profile is shown in Fig. 10. The head and the tail chain segments display density profiles typical for a simple chain in the MC phase.

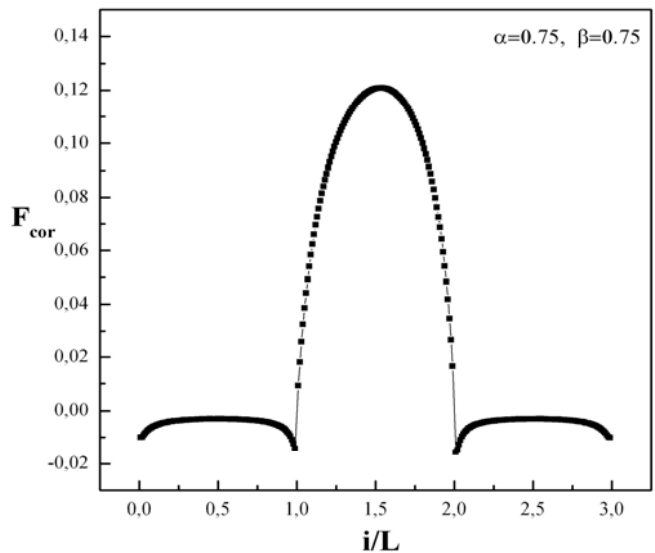

**Figure 11.** Simulation results: Nearest-neighbor correlations $F_{cor}$ in the (MC,CL,MC) state at $\alpha = 0.75$, $\beta = 0.75$.

However, somewhat problematic seems the interpretation of the density profile in the chain segments of the middle section. Instead of being a straight line interpolating between the densities $\rho^{-}_{2,3}, \rho^{+}_{2,3}$ of the low and high density phases, it shows pronounced curvatures near both ends. A detailed analysis shows that these deviations are due to the nearest-neighbor correlations (shown in Fig. 11), which are not negligible.

Similarly to the case $\alpha = \beta = 0.25$, the existence of rather strong cross-correlations is found between equivalent sites belonging to the two branches of the loop, which is quite interesting and unexpected result of our simulations. The spatial dependence of the cross-correlations is shown in Fig. 12.

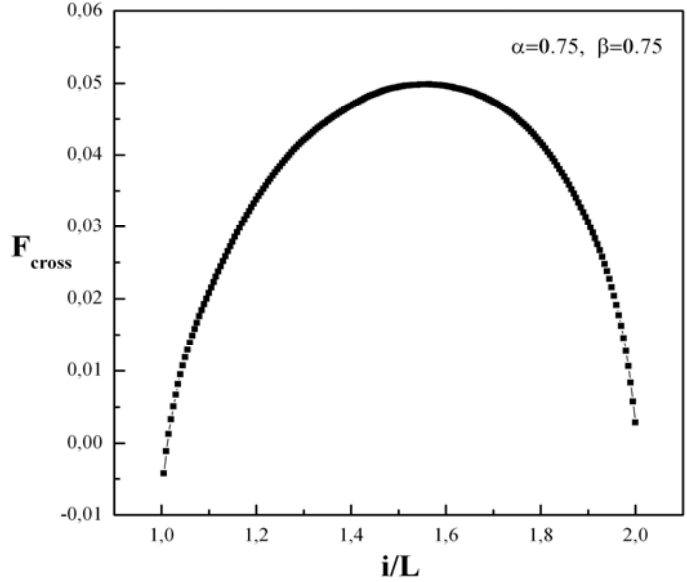

**Figure 12.** Simulation results: Cross-correlations $F_{\text{cross}} = \langle \tau_i^{(1)} \tau_i^{(2)} \rangle - \langle \tau_i^{(1)} \rangle \langle \tau_i^{(2)} \rangle$ between sites $i$ belonging to the different branches of the loop in the (MC, CL, MC) state at $\alpha = 0.75$, $\beta = 0.75$.

## 6. Summary and Conclusions

We have studied the fluctuation properties of the total number of particles $N(t)$ in the totally asymmetric simple exclusion process (TASEP). Computer simulations were performed to obtain and analyze the power spectrum of the total number of particles. The results reveal different types of fluctuation behaviour in the different regions of the phase diagram, thus adding new insight to the very intriguing behavior of the TASEP. In the maximal current phase the fluctuations of $N(t)$ are showing non-trivial behaviour characteristic of SOC. We find that in the whole MC phase the power spectrum of the total number of particles $N(t)$ is characterized by the same exponent $a \cong 1.66$. This result is consistent with the findings in [12]. Across the CL the exponent $a$ changes quite sharply from $a \cong 1.66$ ($\alpha = \beta = 1/2$) to $a \cong 2$ (in the LD and HD phases). The power spectrum of the total number of particles $N(t)$ is very well described by our simple domain wall theory at small values of the injection rate $\alpha$, and the deviation grows for higher values of $\alpha < 1/2$. The latter deviation can be explained by the increasing with $\alpha$ contribution of the fluctuations in the pure low- and high-density phases on both sides of the domain wall.

We have also studied the TASEP on a directed graph with non-trivial topology and open boundaries. The local density profiles, nearest-neighbour correlations along the chain segments, and cross-correlations between equivalent sites belonging to the two branches of the middle section were simulated for values of the parameters $\alpha$ and $\beta$ corresponding to all the phases of a simple chain. The presence of a double-chain middle section leads in some of the cases to expected steady state phase structures, such as (LD,LD,LD) and (HD,HD,HD), which are characterized by short-range correlations appearing in the neighborhood of the bifurcation and merging sites of the network. Otherwise, the properties of the simple-chain segments are close to those expected on the grounds of the approximation which ignores the above mentioned correlations. For example, the reduction of the current by factor of one half in the equivalent branches of the middle section leads to a radical decrease, or increase, of the bulk density in the above cases.

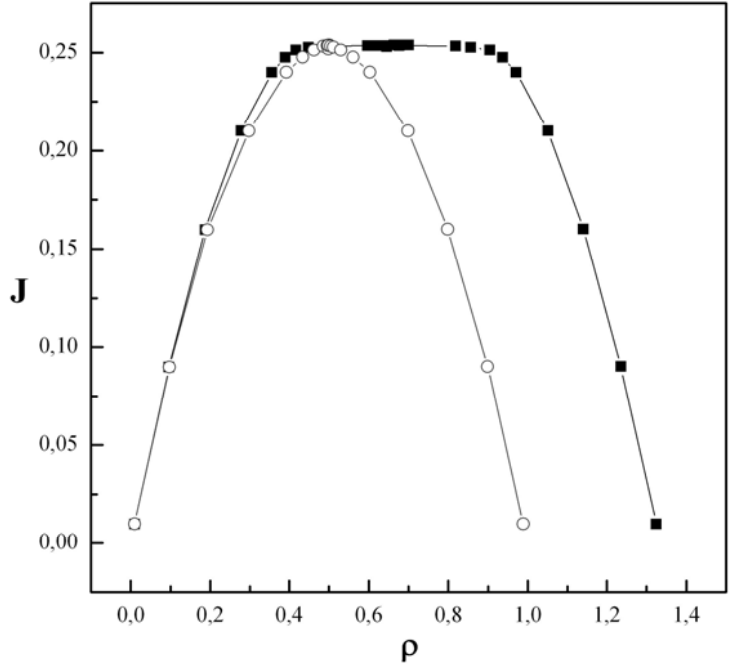

**Figure 13**. Simulation results: The fundamental diagram, current $J$ vs density $\rho$: the solid squares-solid line curve is the result for the system with double chain-section in the middle; the empty circles-dashed line curve is the result for a simple chain.

Rather unexpected are the observed (MC,CL,MC) and what we would call a mixed (LD,CL,HD) and (CL,CL,CL) phase structures. In the former case, which takes place when $\alpha > 1/2$ and $\beta > 1/2$, the bending in opposite directions of the local density profile of the head and tail chains in the maximum-current phase leads to a coexistence of low- and high-density phases in the chain segments of the middle section. The latter case occurs at $\alpha = \beta < 1/2$, when a simple chain is on the coexistence line. For our chain with a double-chain middle section we have found clear evidence of a delocalized domain wall which has different probabilities of being found in the head/tail chains and in the branches of the middle section. No theoretical explanation has been found yet for the significant cross-correlations between the random occupation numbers of equivalent sites belonging to the two branches of the middle section, whenever these branches are in a coexistence phase.

The effect produced by the presence of a double-chain section in our system is very well illustrated when one compares the fundamental diagrams, flow versus density, in our case and in the case of a simple chain, see Fig. 13. The most remarkable features are the appearance of a plateau at the maximum current and the existence of densities greater than unity. To explain the latter feature one has to take into account that, in contrast to the simple chain case, in our network the bulk density happens to be inhomogeneous. The total bulk density is defined as $\rho = \frac{1}{3}\sum_{i=1}^{4}\rho_i$, and the fundamental diagram is calculated as follows: its left-hand half is obtained under fixed $\beta = 0.75$, varying $\alpha \in (0,1)$, and its right-hand half under fixed $\alpha = 0.75$, varying $\beta \in (0,1)$. Remarkably, on going deeper into the maximum-current phase along any of the above paths, the total density of the middle section increases steadily, while the current stays at its maximum value, and the bulk densities in the head and tail segments remain constant too. Since the total density $\rho_2 + \rho_3$ in the two branches of the double-chain section can exceed unity, see Fig. 6, the total bulk density exceeds unity too. The plateau is due to the above mentioned increase of the total density at constant current.

We believe that future investigations on traffic models of complicated single-lane networks are necessary and will reveal new features which have no direct analogs in the simple-chain case. Our preliminary simulations show also that some of the observed correlation effects depend strongly on the length of the head and chain segments.